\documentclass{PoS}

\title{Scale invariance and the electroweak symmetry breaking}

\ShortTitle{Scale invariance and the electroweak symmetry breaking}

\author{\speaker{Archil Kobakhidze}\\
        ARC Centre of Excellence for Particle Physics at the Terascale,\\
        School of Physics, The University of Melbourne, VIC 3010 and \\
        School of Physics, The University of Sydney, NSW 2006, Australia \\ 
        E-mail: \email{archil.kobakhidze@coepp.org.au}}

\abstract{We argue that classical scale invariance provides a technically natural solution to the problem of the radiative stability of the electroweak scale.  Some realistic electroweak scale-invariant models are considered and their possible manifestations at the LHC are briefly discussed.}

\FullConference{36th International Conference on High Energy Physics,\\
		July 4-11, 2012\\
		Melbourne, Australia}

\begin{document}

\section{Motivation}

Scale invariance is a global spacetime symmetry under the rescaling of coordinates and fields according to their canonical dimensions:
\begin{equation}
x^{\mu}\to tx^{\mu}~,~~S(x)\to tS(tx)~,~~V_{\mu}(x)\to tV_{\mu}(tx)~,~~F(x)\to t^{3/2}F(tx)~, 
\label{1}
\end{equation}  
where $S(x),~V_{\mu}(x),~F(x)$ denote scalar, vector and fermion fields, respectively\footnote{Relativistic scale-invariant theories are typically accompanied by a symmetry under special conformal transformations. Together with the $ISO(1,3)$ group of relativistic invariance, scale (dilatation) and special conformal transformations form the 15-parametric conformal group $SO(2,4)$.  In particle physics, however, we are interested in (spontaneously) broken conformal invariance, $SO(2,4)\to ISO(1,3)$, and the only physical manifestation of this breaking is represented by a (pseudo)Goldstone scalar particle, the dilaton, associated with the scale invariance. Therefore, in what follows we will be interested only in scale transformations. }.  In phenomenologically relevant theories, however, scale invariance is an anomalous symmetry, that is, it is an exact symmetry of the classical action, $\sim{\cal O}(\hbar^0)$, and is broken by quantum corrections $\sim {\cal O}(\hbar)$.  An obvious question about scale invariance arises then: Does it matter if it is not actually a symmetry of a full quantum theory?

Firstly, we have a precedent in nature. It is quite remarkable, that most of the mass of the visible matter in the universe carried by protons and neutrons is essentially generated within the scale-invariant QCD through the mechanism of dimensional transmutation. Therefore, it is conceivable to think that the masses of all elementary particles   are also generated in a similar manner. In fact, in technicolour models of electroweak symmetry breaking and alike theories this idea is realised in a non-perturbative fashion. The simplest, QCD-like technicolour models predict (super)heavy composite Higgs boson, which is incompatible the LHC discovery of a Higgs-like particle with mass $m_h\approx 125.5$ GeV announced at this conference \cite{:2012gk}. Dimensional transmutation, however, is a generic phenomenon in clasically scale-invariant theories and is realised also in weekly-coupled, perturbative theories \cite{Coleman:1973jx}. In this talk I will be discussing such perturbative Coleman-Weinberg-type models.

Secondly, scale-invariance may be a low-energy remnant symmetry within a fully conformally-invariant fundamental theory. The best known example of such a theory is string theory, for which conformal invariance, alongside with supersymmetry, is an underlying symmetry. In string theory-motivated phenomenological models, it is more common to assume that conformal invariance is broken at a very high energy scale and supersymmetry survives down to  $\sim$TeV-scale in order to stabilize the electroweak scale against radiative corrections. It seems, however, that the announced mass for the Higgs-like particle is somewhat higher than it is predicted by the simplest supersymmetric extensions of the Standard Model. On the other hand, the radiative stability of electroweak scale can be ensured also by classical scale invariance \cite{bardeen}. Therefore, it is conceivable to consider a version of string theory where supersymmetry is broken at a high energy scale (perhaps not admissible at LHC), while the classical scale invariance survives in the low-energy limit. 

\section{Scale invariance and naturalness}               

There are several theoretical (quantum gravity, unification of forces, electric charge quantization, strong CP problem, etc. ) motivations as well as observational evidences (neutrino masses, dark matter) for new physics that completes the Standard Model at high energies.   Extended theories typically involve different mass scales and the question of whether the electroweak scale is natural arises. A technical aspect of this naturalness problem lies in the fact that, considering the Standard Model as an effective theory valid below a cut-off energy scale $\Lambda$, perturbative corrections to the Higgs boson mass, $m_h$, (and, in fact, to the masses of other Standard Model particles) are of the order of $\Lambda$, i.e., $m_h\sim \Lambda$. Since particle mass is related to spacetime symmetries (an eigenvalue of the quadratic Casimir operator of the Poincar\'e group), an obvious way to solve the problem is to extend relativistic invariance by incorporating a new symmetry that forbids linear dependence of scalar masses on ultraviolet energy scales. The most discussed symmetry that serves this purpose is supersymmetry, another is scale invariance.   

To demonstrate that light scalars are natural (in the above technical sense) within scale-invariant theories, let us consider a simple theory of a single scalar field $S(x)$ which is described the following generating functional:
\begin{equation}
Z_{\Lambda}\left[J_S\right]=\int DS \exp\left\lbrace i\int d^4x  \left({\cal L}_{\Lambda}+J_S S\right) \right \rbrace~,
\label{2}
\end{equation}
where ${\cal L}_{\Lambda}$ is an effective Wilsonian Lagrangian
\begin{equation}
{\cal L}_{\Lambda}=\frac{1}{2}\partial_{\mu}S\partial^{\mu}S-\frac{1}{2}m^2(\Lambda)S^2-\frac{\lambda(\Lambda)}{4}S^4+...
\label{3}
\end{equation}
 where ... denote infinite series of possible terms of mass dimension higher than 4, which are irrelevant at low-energies. In this effective theory scale invariance is badly broken both by the mass term and cut-off $\Lambda$. The effective Wilsonian theory is finite and no further regularization is required. One-loop quantum correction to the $\Lambda$-dependent bare mass in (\ref{3}) can be easily computed:
 \begin{equation}
 m_{\rm R}^{2}(\mu)=m^2(\Lambda)+ \frac{3\lambda(\Lambda)}{16\pi^2}\left[\Lambda^2-m^2(\Lambda)\ln\left( \Lambda^2/\mu^2\right) \right]~. 
\label{4}
 \end{equation}  
Thus, in order to have light scalar, $m_{\rm R}<<\Lambda$, an unnatural fine-tuning between the bare mass and the cut-off scale must be assumed. This is the above-mentioned naturalness (mass hierarchy) problem. 

Suppose now that the effective theory described by (\ref{2}) is embedded in an underlying theory that contains heavy fields/field modes $H(x)$, which is classically scale invariant. 
That is, the generating functional of this `fundamental' theory,
 \begin{equation}
Z\left[J_S, J_H\right]=\int \left[DSDH\right] \exp\left\lbrace i\int d^4x  \left({\cal L}[S,H]+J_S S+J_HH\right) \right \rbrace~,
\label{5}
\end{equation}
is not scale-invariant due to the non-invariance of the functional measure $\left[DSDH\right]$, while the action $\int d^4x {\cal}L[S,H]$ maintains the scale invariance, i.e.:
\begin{equation}
{\cal L}[tS,tH]=t^4{\cal L}[S,H]~.
\label{6}
\end{equation}
Since the bare Wilsonian effective action $\int d^4x {\cal L}_{\Lambda}[S]$ in (\ref{2}) results from integrating out $H$-fields in (\ref{5}), i.e., $\exp\left\lbrace i \int d^4x {\cal L}_{\Lambda}[S] \right\rbrace=\int DH  \exp\left\lbrace i\int d^4x  \left({\cal L}[S,H]\right) \right \rbrace $, the condition
\begin{equation}
m^2_{\rm R}(\Lambda)\equiv m^2(\Lambda)+\frac{3\lambda(\Lambda)}{16\pi^2}\Lambda^2=0~,
\label{7}
\end{equation}  
 is a natural renormalization condition that is forced  due to the classical scale invariance of the underlying theory, see eq. (\ref{6}). Thus we are left only with a mild logarithmic dependence of the scalar mass in (\ref{4}) on the cut-off scale. Other related discussion on the absence of quadratic divergences can be found in \cite{bardeen}, \cite{Foot:2007as}, \cite{Meissner:2007xv}, \cite{Aoki:2012xs}. 

\section{Scale-invariant models}

It is well-known that the scale-invariant version of the Standard Model is not phenomenologically viable since it predicts a very light ($m_h\lesssim 10$ GeV) Higgs boson and, in addition, requires light top quark $m_t\lesssim 40$ GeV. It has been realised for quite sometime  that phenomenologically successful scale-invariant models require extension of the bosonic sector of the Standard Model \cite{Hempfling:1996ht}. There are different physical motivations to do so, e.g., incorporation of neutrino masses, scalar dark matter, new gauge bosons, etc. In ref. \cite{Foot:2010et}  we have observed that incorporation of small cosmological constant within the scale-invariant models necessarily implies that the mass of the dilaton is generated at two-loop level. For the phenomenologically most interesting models with scale-invariance spontaneously broken at TeV-scale, this means that mass of the dilaton can be $\lesssim 10$ GeV. Some realistic electroweak scale-invariant models along these lines have been discussed in \cite{Foot:2011et}. 

The minimal scale-invariant extension of the Standard Model is given by the addition to the Standard Model field content of a singlet scalar field. Demanding cancellation of the cosmological constant, one obtains the following prediction for the dilaton and Higgs mass, respectively           
\begin{eqnarray}
m_{\rm dil}\approx 7-10~{\rm GeV}~, \\
 m_{h}=12^{1/4}m_t\approx 300~{\rm GeV}~.
\label{8}
\end{eqnarray}
If the announced LHC resonance is indeed a Higgs boson, the above minimal model is clearly excluded.  

An interesting class of electroweak scale-invariant theories are those which incorporate type-II see-saw mechanism for neutrino mass generation. These models contain an extra electroweak triplet scalar field $\Delta$. The mass of this particle is predicted to be \cite{Foot:2011et}:
\begin{equation}
m_{\Delta}=\left(2m_t^4-m_h^4/6\right)^{1/4}\approx 190~{\rm GeV}~,
\label{9}
\end{equation} 
for $m_h\approx 125$ GeV. Scale-invariant models with hidden/mirror sector dark matter have also been discussed in \cite{Foot:2011et}. 

\section{Conclusion}

Particle physics models with classical scale invariance are attractive in many respects.  In scale-invariant theories all mass scale have a purely  quantum-mechanical origin.  Very simple models with classical scale invariance are capable of resolving the hierarchy problem without introducing supersymmetry and/or other exotics. Some of the predictions of these theories are  testable at LHC and/or future linear collider.

Phenomenologically the most interesting electroweak scale-invariant models with vanishing cosmological constant generically predict a light dilaton. Realistic scale-invariant models require extended bosonic sector with masses correlated with the masses of the Higgs boson and the top quark. Hopefully, some of the features of scale-invariant theories described in this talk will be observed at LHC.

\vspace{1cm}

I am grateful to Robert Foot, Kristian McDonald and Ray Volkas for collaboration. The work was partially supported by the Australian Research Council.


\vspace{1cm}

\begin{thebibliography}{99}

\bibitem{:2012gk} 
  G.~Aad {\it et al.}  [ATLAS Collaboration],
  Phys.\ Lett.\ B {\bf 716}, 1 (2012)
  [arXiv:1207.7214 [hep-ex]];
  S.~Chatrchyan {\it et al.}  [CMS Collaboration],
  Phys.\ Lett.\ B {\bf 716}, 30 (2012)
  [arXiv:1207.7235 [hep-ex]].

\bibitem{Coleman:1973jx}
  S.~R.~Coleman and E.~J.~Weinberg,
  Phys.\ Rev.\  D {\bf 7}, 1888 (1973).
 
\bibitem{bardeen}
  W.~A.~Bardeen,
  FERMILAB-CONF-95-391-T.

\bibitem{Foot:2007as}
  R.~Foot, A.~Kobakhidze and R.~R.~Volkas,
  Phys.\ Lett.\  B {\bf 655}, 156 (2007)
  [arXiv:0704.1165 [hep-ph]];
  R.~Foot, A.~Kobakhidze, K.~L.~McDonald and R.~R.~Volkas,
  Phys.\ Rev.\  D {\bf 77}, 035006 (2008)
  [arXiv:0709.2750 [hep-ph]].

\bibitem{Meissner:2007xv} 
  K.~A.~Meissner and H.~Nicolai,
  Phys.\ Lett.\ B {\bf 660}, 260 (2008)
  [arXiv:0710.2840 [hep-th]].
  
\bibitem{Aoki:2012xs} 
  H.~Aoki and S.~Iso,
  Phys.\ Rev.\ D {\bf 86}, 013001 (2012)
  [arXiv:1201.0857 [hep-ph]].
  

  \bibitem{Hempfling:1996ht}
  R.~Hempfling,
  Phys.\ Lett.\  B {\bf 379}, 153 (1996)
  [arXiv:hep-ph/9604278]; 
  W.~F.~Chang, J.~N.~Ng and J.~M.~S.~Wu,
  Phys.\ Rev.\  D {\bf 75}, 115016 (2007)
  [arXiv:hep-ph/0701254];
  R.~Foot, A.~Kobakhidze, K.~L.~McDonald and R.~R.~Volkas,
  Phys.\ Rev.\  D {\bf 76}, 075014 (2007)
  [arXiv:0706.1829 [hep-ph]];
  T.~Hambye and M.~H.~G.~Tytgat,
  Phys.\ Lett.\  B {\bf 659}, 651 (2008)
  [arXiv:0707.0633 [hep-ph]]; 
  S.~Iso, N.~Okada and Y.~Orikasa,
  Phys.\ Lett.\  B {\bf 676}, 81 (2009)
  [arXiv:0902.4050 [hep-ph]];
  M.~Holthausen, M.~Lindner and M.~A.~Schmidt,
  Phys.\ Rev.\  D {\bf 82}, 055002 (2010)
  [arXiv:0911.0710 [hep-ph]];
  R.~Foot, A.~Kobakhidze and R.~R.~Volkas,
  Phys.\ Rev.\  D {\bf 82}, 035005 (2010)
  [arXiv:1006.0131 [hep-ph]].
  L.~Alexander-Nunneley and A.~Pilaftsis,
  JHEP {\bf 1009}, 021 (2010);
  [arXiv:1006.5916 [hep-ph]]. 
  A.~Latosinski, K.~A.~Meissner and H.~Nicolai,
  arXiv:1010.5417 [hep-ph].
  
\bibitem{Foot:2010et} 
  R.~Foot, A.~Kobakhidze and R.~R.~Volkas,
  Phys.\ Rev.\ D {\bf 84}, 075010 (2011)
  [arXiv:1012.4848 [hep-ph]].
  
\bibitem{Foot:2011et} 
  R.~Foot and A.~Kobakhidze,
  arXiv:1112.0607 [hep-ph].







\end{thebibliography}
\end{document}